\documentclass[%
 reprint,
 amsmath,amssymb,
 aps,
]{revtex4-2}
\usepackage{mathtools}
\usepackage{physics}
\usepackage{braket}
\usepackage{here}
\usepackage{graphicx}
\usepackage{dcolumn}
\usepackage{bm}
\usepackage{newtxtext,newtxmath}
\usepackage{hyperref}
\hypersetup{
setpagesize=false,
bookmarksnumbered=false,%
bookmarksopen=false,%
colorlinks=true,%
linkcolor=blue,
citecolor=blue,
urlcolor=blue,
}
\usepackage[normalem]{ulem}
\usepackage{xcolor}
\usepackage{lineno}

\usepackage{ragged2e}

\newcommand{\br}{{\mathbf{r}}}

\newcommand{\zetatr}{{\zeta_{\rm tr}}}
\newcommand{\zetarot}{{\zeta_{\rm rot}}}

\newcommand{\Delete}[1]{}

\newcommand{\Note}[1]{}

\begin{document}

\title{Chemotaxis-induced clustering of self-propelled rods}

\author{Shun Sakurai}
\author{Nariya Uchida}%
 \email{nariya.uchida@tohoku.ac.jp}
\affiliation{%
 Department of Physics, Tohoku University, Sendai 980-8578, Japan
}%

\date{\today}

\begin{abstract}
We investigated how active rod-like particles with chemotaxis and orientational interactions, such as rod-shaped bacteria or chemically interacting Janus rods, undergo cluster formation. We constructed a self-propelled rod model incorporating two types of chemotaxis—rotational and translational—and numerically examined the collective behavior of the particle system. Both rotational and translational chemotaxis promote particle aggregation through distinct mechanisms. 
Rotational chemotaxis has a minimal impact on extending the range of noise intensities over which clustering is observed, yet it enhances cluster growth, leading to larger cluster sizes. Conversely, translational chemotaxis both extends this range to higher noise intensities and increases the maximum cluster size.
\end{abstract}


\maketitle


\section{\label{sec:introduction}Introduction}

Chemotaxis is a key mechanism governing the motion of microorganisms. Eukaryotic cells migrate toward regions of higher chemical concentration by sensing spatial gradients of attractants~\cite{parent1999cell}. In contrast, bacteria detect temporal changes in concentration and adjust their tumbling probability during run-and-tumble motion, allowing them to bias their movement toward regions of higher concentration~\cite{berg1975chemotaxis}. 
It is known that bacteria can release chemical attractants themselves and attract others via autochemotaxis, thereby forming clusters with diverse morphologies~\cite{budrene1991complex,budrene1995dynamics,mittal2003motility,berg2004coli}.
The formation of such clusters has been theoretically modeled using reaction-diffusion equations, incorporating variables such as bacterial density and the concentration field of chemical attractants~\cite{aotani2010model,brenner1998physical,keller1970initiation}.

On the other hand, many bacterial species possess rod-like shapes and exhibit collective behavior characterized by orientational alignment. Rod-shaped particles with such alignment interactions are known as self-propelled rods~\cite{ginelli2010large}. The self-propelled rod model is a modification of the classical Vicsek model~\cite{vicsek1995novel,chate2008collective},
incorporating alignment interactions in which two particles tend to align parallel if their relative angle is acute, and antiparallel if it is obtuse. Recent experiments using elongated mutant strains of bacteria ~\cite{nishiguchi2017long} 
have demonstrated that bacterial populations can exhibit long-range orientational 
order and reproduce the statistical properties predicted by self-propelled rod models.

In addition to biological systems, synthetic microswimmers such as Janus particles provide experimentally controllable platforms to study chemotactic behavior.
Janus particles have two distinct faces, one of which selectively catalyzes chemical reactions to generate concentration gradients. 
These gradients induce fluid flows that drive their 
self-propulsion~\cite{howse2007self,walther2013janus}. 
Rod-shaped Janus particles, or Janus rods~\cite{paxton2004catalytic,paxton2005motility}, 
are known to form locally aligned clusters~\cite{vutukuri2016dynamic}.

Several theoretical models have been proposed to describe chemotaxis 
at the particle level. 
These include active Brownian particle models~\cite{pohl2014dynamic,stark2018artificial}, 
{models incorporating catalytic reactions in Janus particles~\cite{saha2014clusters,tucci2024nonreciprocal},
rotational chemotaxis in chemically interacting rotors~\cite{liebchen2016pattern}, 
and a phase field approach~\cite{paspunurwar2024dynamic}.
These models predict a variety of self-organized patterns, including dynamic 
clusters, bands, and asters.
A comprehensive review of synthetic chemotaxis and collective behavior in active matter is provided in~\cite{liebchen2018synthetic}.}
Hydrodynamic interactions have additionally been incorporated into 
Fokker-Planck-type equations to predict band-like aggregation patterns~\cite{lushi2012collective,lushi2018nonlinear}.
More recently, frameworks integrating chemotactic responses with velocity or polar alignment mechanisms have also been introduced~\cite{di2020hybrid, ziepke2022multi}, reporting outcomes ranging from static point-like aggregations to dynamic droplets, streams, vortices, and spiral signaling waves.

However, the interplay of autochemotaxis with nematic alignment—characteristic of self-propelled rods—remains largely unexplored. Because chemotaxis involves particle responses to scalar concentration gradients whereas nematic alignment arises from vectorial coupling between particle orientations, their combined effects operate through fundamentally different mechanisms and cannot be captured by simply superimposing known results from each process. A theoretical framework that treats nematic alignment alongside chemotactic mechanisms in a unified manner is thus essential for understanding systems where both effects are present—such as in populations of rod-shaped bacteria or Janus rods that produce and respond to chemical cues while aligning locally.

In this study, we incorporate chemotaxis into the self-propelled rod model to investigate how chemotaxis affects clustering in self-propelled rods.
We analyze the resulting collective behavior using numerical simulations.

We consider two types of chemotactic responses of self-propelled rods that produce chemoattractants: translational chemotaxis, which induces translational drift along the concentration gradient, and rotational chemotaxis, which reorients the particle toward the concentration gradient~\cite{pohl2014dynamic,liebchen2016pattern}. 

The structure of the paper is as follows. In Section II, we set up the model. In Section III, we describe the numerical calculation methods. In Section IV, we present the results. Finally, in Section V, we discuss our findings in the context of related experimental and theoretical studies and conclude.

\section{\label{sec:model}Model}

We set up a two-dimensional model of self-propelled rods with chemotactic interactions mediated by the concentration field \( c(\mathbf{r}, t) \) of a chemoattractant.
We start from the discrete-time evolution equations:
\begin{align}
\mathbf{r}^{t+\Delta t}_j &= \mathbf{r}^t_j 
+ v_0\Delta t \, \mathbf{e} \left( \theta^{t+\Delta t}_j \right)
+ \zetatr \nabla c \left( \mathbf{r}_j^t \right) \Delta t,
\label{eq:rt}
\\
\theta^{t+\Delta t}_j &= \arg 
\left[
\sum_{k; |\mathbf{r}_j - \mathbf{r}_k| < d_0} \mathrm{sign} 
\left\{ \cos \left( \theta^t_k - \theta^t_j \right) \right\}
 \exp \left( i\theta^t_k \right)
\right] + \eta_j^t
\nonumber\\
&\quad + 
\zetarot \, 
\partial_\theta 
\mathbf{e} \left( \theta_j^t \right) 
\cdot \nabla c \left( \mathbf{r}_j^t \right) \Delta t
+ \overline{\eta}^t_j
\label{eq:thetat}
\end{align}
where \( \mathbf{r}_j^t \) and \( \theta_j^t \) are the position and orientation of the \( j \)-th particle at time \( t \), respectively;  
\( \Delta t \) is the time step;  
\( \mathbf{e}(\theta) = (\cos\theta, \sin \theta) \) is the orientation vector;  
\( v_0 \) is the self-propulsion speed;  
\( d_0 \) is the interaction radius for alignment;  
\( \zetatr \) and \( \zetarot \) are the translational and rotational chemotactic susceptibilities, respectively;
$\eta_j^t$, and $\overline{\eta}_j^t$ are independent angular noise terms. 
The angular noise terms $\eta_j^t$ and  $\overline{\eta}_j^t$ are drawn 
from the Gaussian distribution  
$P(\eta) = \frac{1}{\sqrt{2\pi}\sigma} \exp\left(-\frac{\eta^2}{2\sigma^2}\right)$.
$\eta_j^t$ is a collisional noise term that is non-zero only when other particles are present within the interaction radius $d_0$, and is zero otherwise. $\overline{\eta}_j^t$ is a self-diffusive noise term that is applied according to a Poisson process with unit rate.
The sum in \eqref{eq:thetat} includes the particle itself (\(k=j\)).
At each time step, the orientation is first updated according to
Eq.~\eqref{eq:thetat}, and the particle position is then updated
according to Eq.~\eqref{eq:rt} using the updated orientation
$\theta_j^{t+\Delta t}$.

Each particle secretes chemoattractant at a rate \( a_1 \), while the substance decays at rate \( a_2 \). The chemoattractant concentration field $c(\br,t)$
evolves according to the equation:
\begin{align}
\frac{\partial c}{\partial t} = D_c\nabla^2 c + a_1 \sum_j \delta(\mathbf{r} - \mathbf{r}_j^t) - a_2 c,
\label{eq:ct}
\end{align}
where \( D_c \) is the diffusion constant.

\section{\label{sec:methods} NUMERICAL SIMULATION METHODS}

We describe the numerical simulation methodology used in this study. 
We consider an $L\times L$ square domain with periodic boundary conditions, with $L=100$.
The particle self-propulsion speed is $v_0 = 1.0$ and the interaction radius is $d_0 = 0.5$. At each time step, each particle interacts with 
all particles within its interaction radius, following Eq.\ref{eq:thetat}. 
Orientational noise $\eta_j^t$ is added during this alignment process. 
To account for chemotactic drift and prevent particles from jumping past another particle's interaction radius—which would cause unphysically missed interactions—the time step size is fixed at $\Delta t = 0.01$. Unless otherwise stated, the simulations are run up to $t=1500$. The self-diffusive noise term $\overline{\eta}_j^t$ is modeled as a Poisson process event occurring at a rate of once per unit time, taking place with a probability $p = 1 - e^{-\Delta t}$ within a time step $\Delta t$. The total number of particles is $N$, and the particle number density is defined as $\rho = N/L^2$. Particles are initially distributed uniformly at random in the simulation domain, and their orientations are independently drawn from a uniform distribution over $[-\pi,\pi)$.

The chemical concentration field is discretized on a $250 \times 250$ spatial grid with a cell size $d_{c} = 0.4$, subject to periodic boundary conditions. The time evolution of the concentration field is numerically integrated using the explicit Euler method. Each particle perceives the concentration gradient based on the concentration differences among the four neighboring cells (top, bottom, left, and right) relative to its host grid cell, modulating its speed and orientation through chemotaxis. Furthermore, the chemical emitted by each particle is added directly to its host grid cell as a concentration increment of $\Delta ta_1/d_c^2$ per particle at each time step. The chemical field parameters are set to a diffusion coefficient of $D_c = 0.4$, a particle production rate of $a_1 = 0.05$, and a degradation rate of $a_2 = 0.01$.
The initial chemical concentration $c_0$ for the field $c$ is set to the equilibrium concentration of the homogeneous mean-density state $c_0 = \rho a_1 / a_2$, as determined by Eq.\ref{eq:ct}.
The numerical timestep $\Delta t = 0.01$ satisfies the stability condition for the diffusion equation, as it remains below the upper threshold limit $d_{\mathrm{c}}^2 / 4 D_{\mathrm{c}}$.
We performed calculations for particle number density $\rho$ ranging from $0.1$ to $1.0$ (i.e., $N = 1000$ to $10,000$) and noise intensity $\sigma^2$ ranging from $0$ to $0.24$.
All quantities are expressed in dimensionless simulation units.

\section{\label{sec:results} RESULTS}

\begin{figure*}[htbp]
    {\centering \includegraphics[width=175mm]{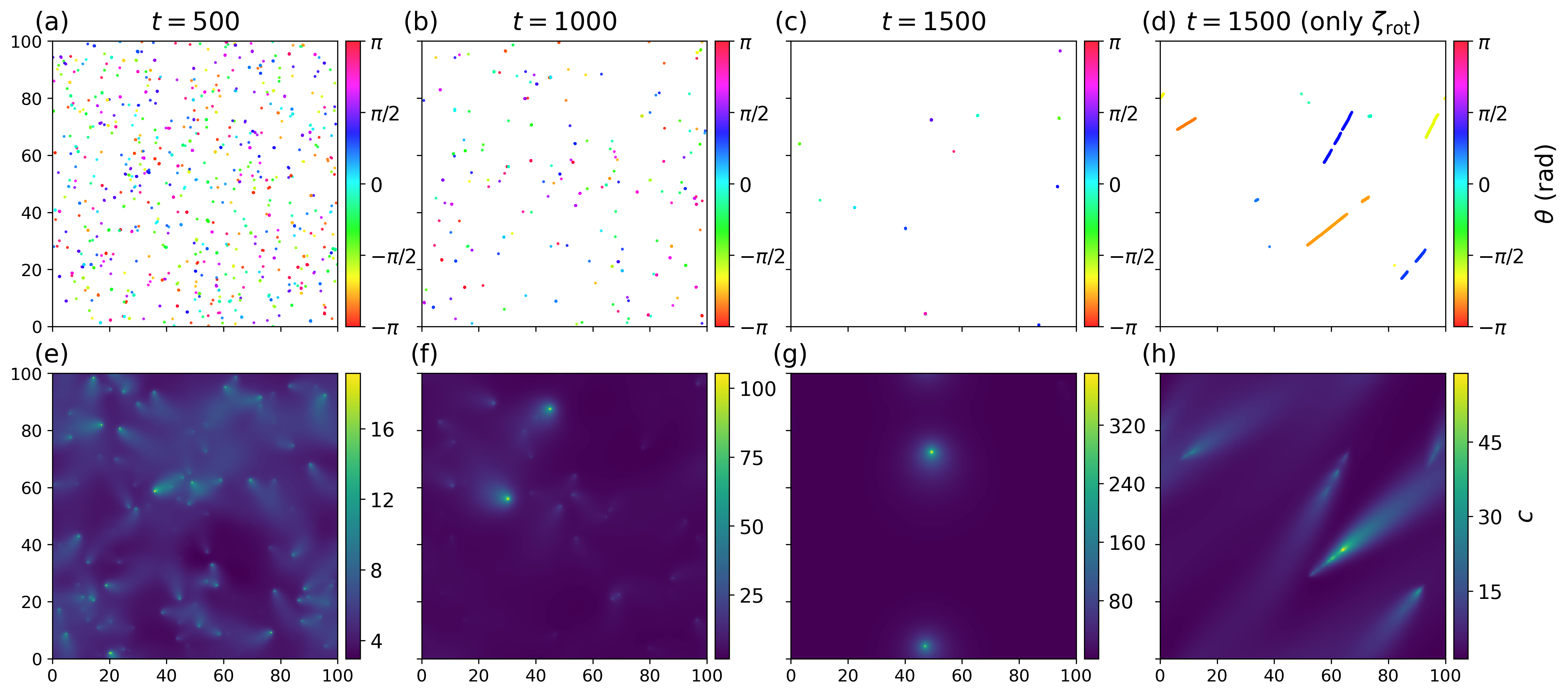}\par}
    \caption{
    Snapshots of the particle and chemical concentration fields for $(\zetarot, \zetatr, \rho, \sigma^2)=(0.05, 0.05, 1.0, 0.06)$. (a)–(c) Snapshots of particle positions at times $t = 500$, $1000$, and $1500$. Color scale represents the particle orientation angle $\theta \in [-\pi, \pi]$ (rad). (e)–(g) Corresponding spatial distribution of chemical concentration $c$ at $t = 500$, $1000$, and $1500$. Color bars on the right indicate the concentration level $c$. (d), (h) are snapshots of particles and chemical concentration at $t = 1500$ for $(\zeta_{\text{rot}}, \zeta_{\text{tr}}, \rho, \sigma^2) = (0.05, 0.0, 1.0, 0.0)$.
    }
    \label{fig:snapshots}
\end{figure*}

Fig.\ref{fig:snapshots} presents an overview of the simulation results. When chemotaxis exceeds a certain threshold, particles accumulate over time, resulting in aggregation. At $t = 1500$, the system segregates into a few distinct clusters. The bottom panel of Fig.\ref{fig:snapshots} shows the spatial distribution of the chemical concentration. 
Chemotaxis induces local regions of elevated particle density; because higher particle density increases local chemical production, particles are further attracted to these areas. Through this self-reinforcing mechanism, particles continuously gather around existing aggregates, ultimately leading to the formation of a few large clusters. At $t = 1500$, two high-concentration chemical spots are visible, each corresponding to one of the large clusters.

On the other hand, when only rotational chemotaxis is present (i.e., in the absence of translational chemotaxis), clusters concentrated at a single point are not formed. Instead, linear chains of particles develop along their direction of motion. This is likely because the chemical produced by the particles leaves a trail along their trajectories, and 
rotational chemotaxis steers other particles toward the chemical trail, after which nematic interactions align them with the moving cluster.
Note that in the absence of nematic interactions, point-like clusters concentrated at a single point are formed regardless of whether only translational or rotational chemotaxis is present.

A cluster $C$ is defined as a connected component of the resulting particle network, representing a maximal mutually reachable group of particles. 
Two particles are directly bonded if their distance satisfies $r_{ij} < r_{\mathrm{cut}}$, taking into account periodic boundary conditions. 
We set $r_{\mathrm{cut}}=d_0=0.5$.
In the analysis of individual cluster properties,
we restrict ourselves to clusters satisfying
$N_C \geq N_{\min}$, where $N_{\min}=10$.
No such restriction is imposed when calculating $f_{\max}$.

\begin{figure*}[htbp]
{\centering
\includegraphics[width=175mm]{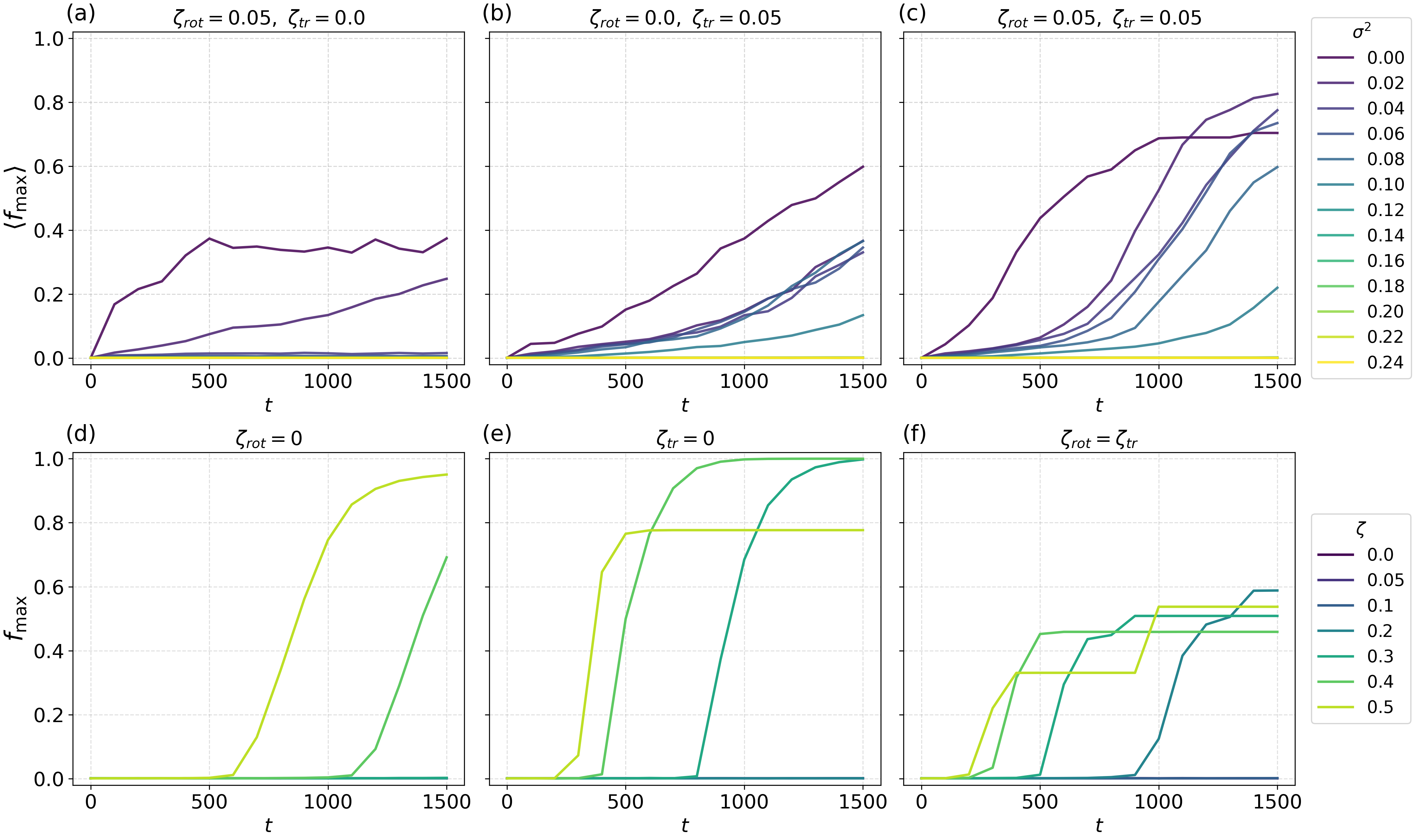}\par}
\caption{\justifying
(a)–(c) Time evolution of the ensemble-averaged maximum cluster fraction $\langle f_{\max} \rangle$, defined as the ratio of the maximum cluster size to the total number of particles, averaged over 10 independent simulation runs for $(\zeta_{\text{rot}}, \zeta_{\text{tr}}) = (0.05, 0.0)$, $(0.0, 0.05)$, and $(0.05, 0.05)$, respectively, with $\rho = 1.0$. (d)–(f) Time evolution of $f_{\max}$ in the absence of nematic interactions for $\zeta_{\text{rot}} = 0.0$, $\zeta_{\text{tr}} = 0.0$, and $\zeta_{\text{rot}} = \zeta_{\text{tr}}$, respectively, with $(\rho, \sigma^2) = (1.0, 0.0)$. The legend values correspond to $\zeta_{\text{tr}}$ in (d), $\zeta_{\text{rot}}$ in (e), and $\zeta_{\text{rot}} = \zeta_{\text{tr}}$ in (f), respectively.
}
\label{fig:clustersize}
\end{figure*}

The clustering behavior of the particles is shown in Fig.\ref{fig:clustersize}. Panels (a)–(c) depict the time evolution of clustering for different chemotactic coefficients and noise intensities at $\rho = 1.0$. 
Here, $f_{\max}=N_{\max}/N$ is defined as the ratio
of the maximum cluster size $N_{\max}$ to the total
number of particles $N$.
Panels (a)--(c) show its ensemble average
$\langle f_{\max}\rangle$ over 10 independent runs.
When only rotational chemotaxis is present (a), clustering is observed only at low noise intensities, namely $\sigma^2 = 0.00$ and $\sigma^2 = 0.02$; for larger noise levels, 
$\langle f_{\max}\rangle$ remains close to zero. In contrast, when only translational chemotaxis or both translational and rotational chemotaxis are present, clustering occurs up to $\sigma^2 = 0.10$. For comparison, we calculated the time evolution of $f_{\max}$ in the absence of nematic interactions, i.e., with chemotaxis alone [(d)–(f)], for $(\rho, \sigma^2) = (1.0, 0.0)$. Specifically, the calculation was performed by replacing the alignment term in Eq.\ref{eq:thetat} with $\theta_j^t$ and omitting the collision noise term $\eta_j^t$. Within the observation window up to $t = 1500$, a sharp increase in $f_{\max}$ becomes apparent between $\zeta_{\text{tr}} = 0.3$ and $0.4$ for translational chemotaxis and between $\zeta_{\text{rot}} = 0.2$ and $0.3$ for rotational chemotaxis. 
When the simulations are extended to $t=15000$,
$f_{\max}$ reaches unity for $\zeta_{\text{tr}}=0.2$
in the presence of translational chemotaxis alone.
However, no significant increase is observed for
$\zeta_{\text{tr}}=0.1$ or $\zeta_{\text{rot}}=0.2$
in the respective cases.
Thus, within the observation window of $t\leq1500$,
clustering occurs at $\zeta=0.05$ in the presence
of nematic interactions for either type of chemotaxis,
whereas larger chemotactic coefficients are needed
in their absence.
Even when both types of chemotaxis are present,
clustering in the absence of nematic interactions
is observed at
$\zeta_{\text{rot}}=\zeta_{\text{tr}}=0.2$
within this time window.

\begin{figure*}[htbp]
{\centering
\includegraphics[width=160mm]{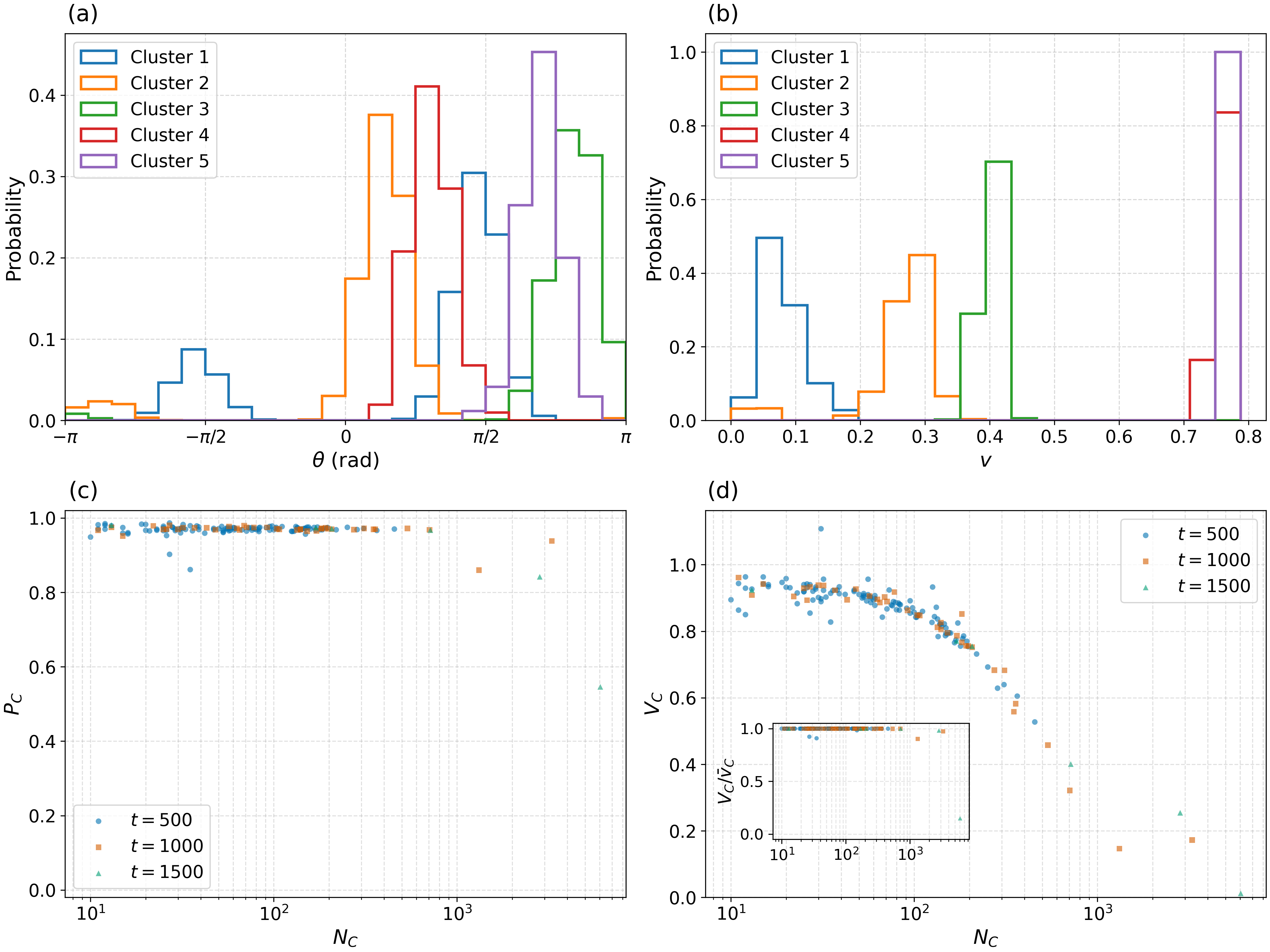}\par}
\caption{\justifying
(a), (b) Distributions of orientation angle and speed of constituent particles in the five largest clusters at $t = 1500$ for $(\zeta_{\text{rot}}, \zeta_{\text{tr}}, \rho, \sigma^2) = (0.05, 0.05, 1.0, 0.06)$. The sizes of Clusters 1, 2, 3, 4, and 5 are 6050, 2827, 715, 207, and 170, respectively. (c), (d) Polar order parameter and cluster speed as a function of cluster size at $t = 500, 1000,$ and $1500$ for $(\zeta_{\text{rot}}, \zeta_{\text{tr}}, \rho, \sigma^2) = (0.05, 0.05, 1.0, 0.06)$. The inset shows the ratio of the cluster speed $V_C$
to the mean particle speed $\bar{v}_C$.
}
\label{fig:clusteranalysis}
\end{figure*}

We now examine the specific structures of the formed clusters in detail. Fig.\ref{fig:snapshots}(c) shows the formation of clusters of various sizes. Here, we focus on the clusters formed at $(\zeta_{\text{rot}}, \zeta_{\text{tr}}, \rho, \sigma^2) = (0.05, 0.05, 1.0, 0.06)$, which corresponds to the same parameter set as in this snapshot. Fig.\ref{fig:clusteranalysis}(a) and (b) show the distributions of orientation and speed of the constituent particles within the five largest clusters. The orientation distribution of constituent particles in Cluster 1, the largest cluster, exhibits peaks near $\pi/2$ and $-\pi/2$, indicating that the particles are aligned in antiparallel directions. In contrast, smaller clusters display a single prominent peak with few antiparallel particles. Furthermore, the particle speed distribution reveals that the average speed decreases as the cluster size (number of particles) increases. Fig.\ref{fig:clusteranalysis}(c) and (d) plot the polar order parameter $ P_C = \sqrt{\langle \cos\theta_j^t \rangle_C^2 + \langle \sin\theta_j^t \rangle_C^2} $ and cluster speed $V_C = |\langle \mathbf{v}_j^t \rangle_C|$ for each cluster at $t = 500$, $1000$, and $1500$, 
where $\mathbf{v}_j^t$ is the velocity of the $j$-th particle estimated
from its displacement between consecutive stored frames separated by
$\Delta T=1$, taking periodic boundary conditions into account, and the
averages are taken over particles in the cluster labeled by $C$.
For $t=500$ and $1000$, the displacement to the subsequent stored frame
is used, whereas for $t=1500$, the displacement from the preceding
stored frame is used.
As shown in Fig.\ref{fig:clusteranalysis}(c), except for a few giant clusters, the polar order parameter is close to $1$, indicating that the constituent particles are oriented in nearly the same direction. Although not shown in the figure, the nematic order parameter $Q_C = \sqrt{\langle \cos(2\theta_j^t) \rangle_C^2 + \langle \sin(2\theta_j^t) \rangle_C^2}$ remains approximately $0.9$ over a wide range of cluster sizes, including the largest clusters. 
Thus, the decrease in $P_C$ in the largest clusters reflects the emergence of antiparallel orientations rather than a loss of orientational order. 
The cluster speed decreases from the self-propulsion speed $v_0 = 1.0$ with increasing cluster size. Notably, the cluster speed is lower than $v_0$ even when the polar order parameter is close to $1$; this demonstrates that the speed of individual constituent particles is reduced by translational chemotaxis, as seen in Fig.\ref{fig:clusteranalysis}(b). As seen in the inset of Fig.\ref{fig:clusteranalysis}(d), except for a single cluster, the cluster speed $V_C$ is nearly equal to the mean particle speed $\bar{v}_C = \langle \vert{}{\bf v}_j^t \vert{} \rangle_C$.

\begin{figure}[htbp]
{\centering
\includegraphics[width=85mm]{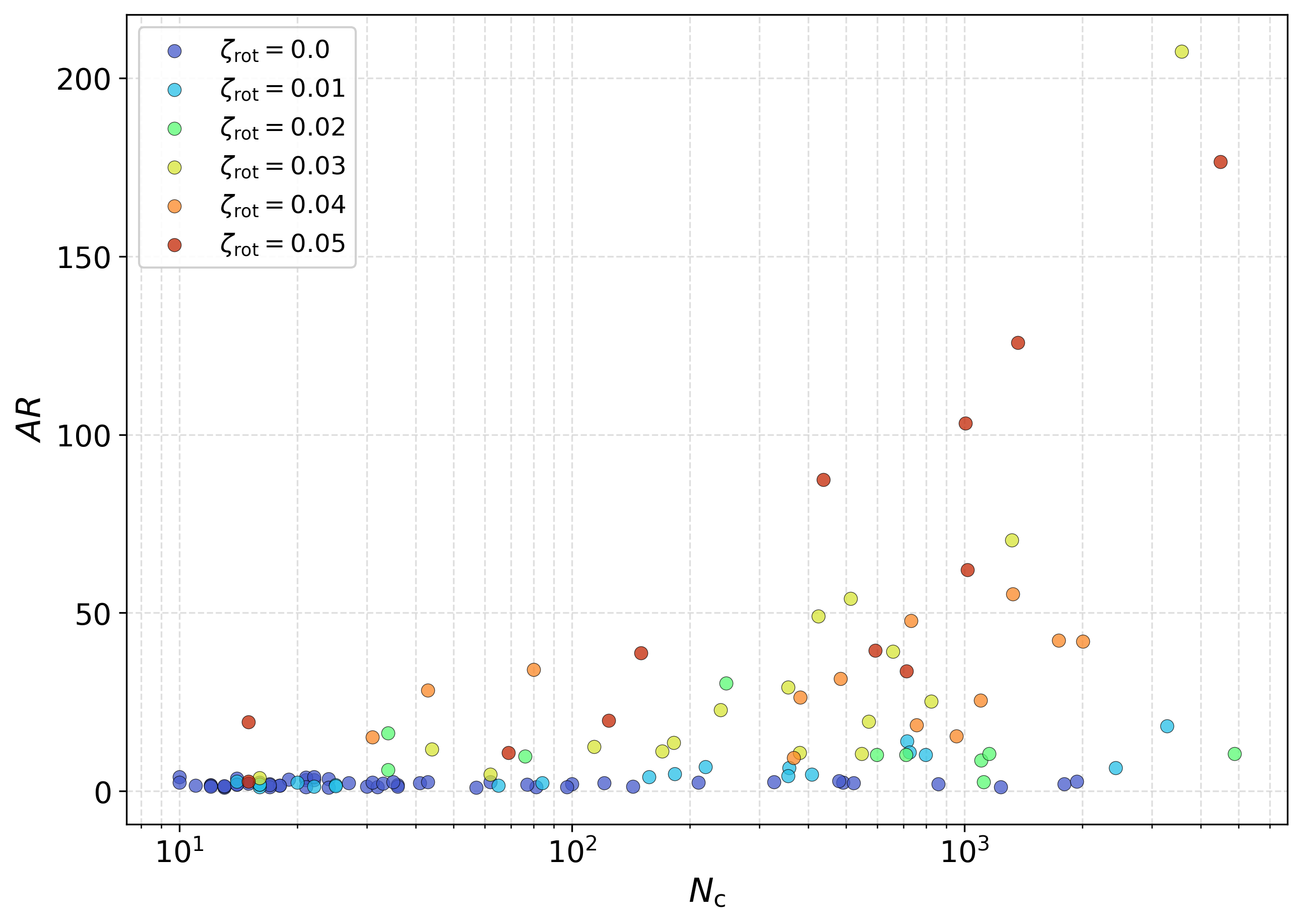}\par}
\caption{\justifying
Aspect ratio and size of individual clusters for various values of $\zeta_{\text{rot}}$ at $(\zeta_{\text{tr}}, \rho, \sigma^2) = (0.0, 1.0, 0.0)$ and $t=1500$.
}
\label{fig:aspectratio}
\end{figure}

In the presence of rotational chemotaxis alone (without translational chemotaxis), at $\sigma^2 = 0.0$ where clustering occurs, the polar order parameter remains close to $1$ even for large cluster sizes, and the cluster speed is nearly equal to the self-propulsion speed of $1.0$. Fig.\ref{fig:aspectratio} shows the aspect ratio ($\mathrm{AR}$) of individual clusters for various values of $\zeta_{\text{rot}}$ at $(\zeta_{\text{tr}}, \rho, \sigma^2) = (0.0, 1.0, 0.0)$. 
The aspect ratio $\mathrm{AR}$ of a cluster is evaluated via the eigenvalues $\lambda_1$ and $\lambda_2$ ($\lambda_1 \ge \lambda_2$) of its gyration tensor $\mathbf{S}$, defined as $\mathrm{AR} = \sqrt{\lambda_1 / \lambda_2}$. Here, the components of $\mathbf{S}$ are given by $S_{\alpha\beta} = \frac{1}{N_C} \sum_{i\in C} (r_{i,\alpha} - r_{C,\alpha})(r_{i,\beta} - r_{C,\beta})$ ($\alpha, \beta \in \{x, y\}$), where $r_{i,\alpha}$ and $r_{C,\alpha}$ denote the $\alpha$-component of the $i$-th particle's position and the cluster's center-of-mass position $\mathbf{r}_C$, both evaluated using unwrapped coordinates to account for periodic boundary conditions.
It reveals that $\mathrm{AR}$ increases with increasing $\zeta_{\text{rot}}$, leading to elongated rod-like shapes. In these rod-like clusters, the direction of the major axis coincides with the orientation direction of the constituent particles, and the clusters move parallel to their major axis.

\begin{figure*}[htbp]
{\centering
\includegraphics[width=140mm]{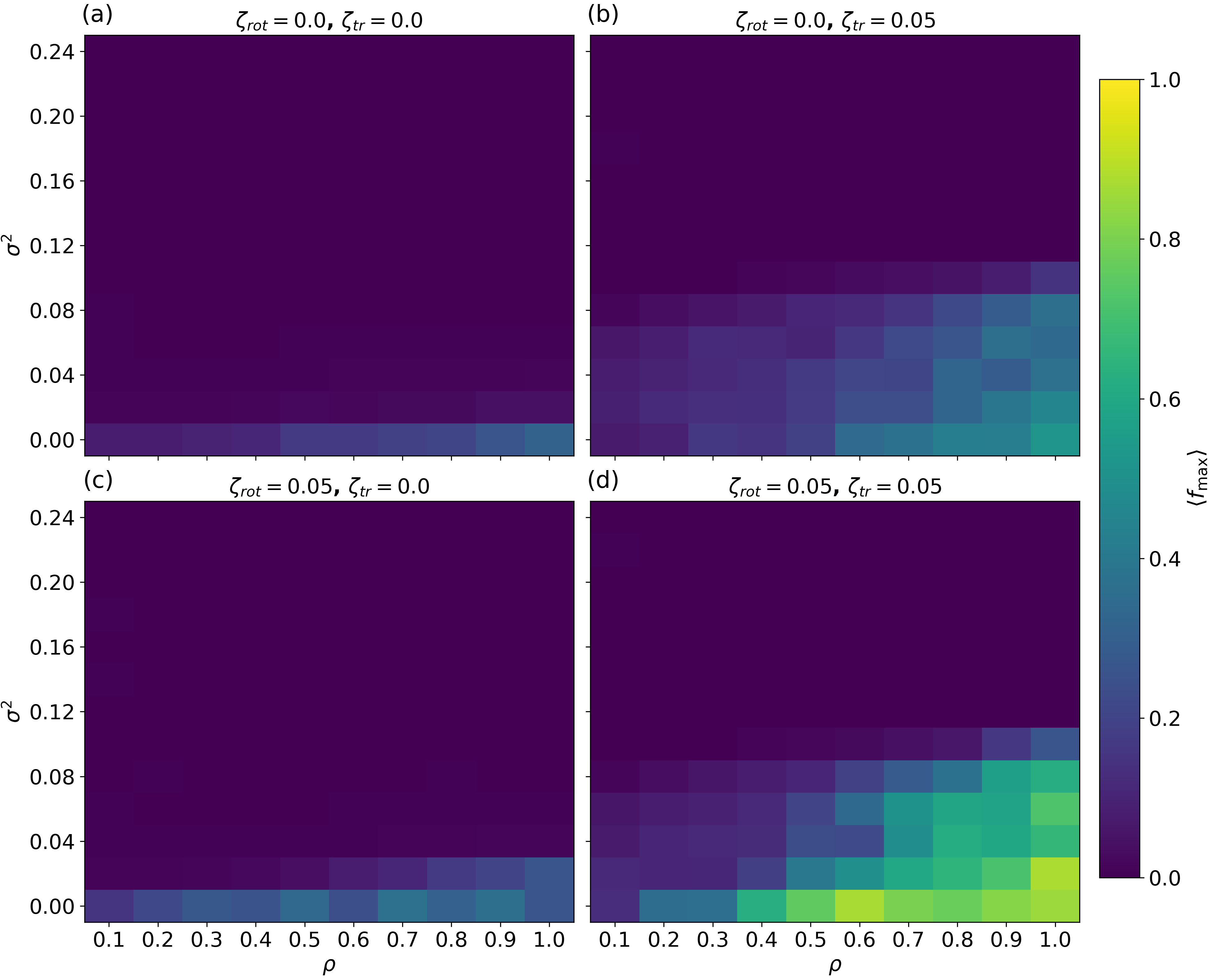}\par}
\caption{\justifying The maximum cluster fraction $\langle f_{\max}\rangle$ (the ratio of the maximum cluster size to the total number of particles) at different chemotactic coefficients for various values of $(\rho, \sigma^2)$ at $t=1500$: (a) $(\zeta_{\text{rot}}, \zeta_{\text{tr}}) = (0.0, 0.0)$, (b) $(\zeta_{\text{rot}}, \zeta_{\text{tr}}) = (0.0, 0.05)$, (c) $(\zeta_{\text{rot}}, \zeta_{\text{tr}}) = (0.05, 0.0)$, and (d) $(\zeta_{\text{rot}}, \zeta_{\text{tr}}) = (0.05, 0.05)$. Data are ensemble-averaged over 10 independent runs for $\sigma^2 \in [0.0, 0.04]$ in (a) and (c), and for $\sigma^2 \in [0.0, 0.12]$ in (b) and (d), while a single run is shown for all other values of $\sigma^2$.
}
\label{fig:phasediagram}
\end{figure*}

Using the maximum cluster fraction at $t=1500$
for each $(\rho,\sigma^2)$, we constructed color maps
indicating the degree of clustering within the simulated
time window (Fig.~\ref{fig:phasediagram}).
We averaged the data over 10 independent runs in and
around the parameter regions exhibiting appreciable
clustering, while using single-run data in the high-noise regions (see the figure caption for details).

In the absence of chemotaxis (a), cluster formation is primarily restricted to $\sigma^2=0.0$. 
When only rotational chemotaxis is present (c), clustering at $\sigma^2=0.0$ is generally enhanced, and cluster formation is also facilitated at $\sigma^2=0.02$ in the high-density region (large $\rho$). 
In contrast, when only translational chemotaxis is present (b), clustering is promoted even at relatively high noise intensities up to $\sigma^2=0.10$.
When both types of chemotaxis are present (d),
the range of noise intensities favoring cluster formation is similar to that for translational chemotaxis alone, but large-scale clusters are more prominent, with $\langle f_{\max}\rangle$ exceeding $0.6$ for several parameter sets.
These findings reveal that rotational chemotaxis has a minor effect on expanding the high-$\langle f_{\max}\rangle$ domain but generally increases the magnitude of $\langle f_{\max}\rangle$.
Meanwhile, translational chemotaxis both expands the high-$\langle f_{\max}\rangle$ domain and boosts its value, with their combination producing a further enhancement of $\langle f_{\max}\rangle$, as seen in (d).

\section{\label{sec:discussion} DISCUSSION}

Several studies have investigated the collective motion of autochemotactic self-propelled particles. These include models incorporating rotational chemotaxis alongside hydrodynamic interactions where particle motion induces fluid flow in the chemical solution~\cite{lushi2018nonlinear}, models relying on translational chemotaxis without a constant self-propulsion speed and introducing velocity-alignment forces based on velocity differences with neighboring particles~\cite{di2020hybrid}, and models featuring rotational chemotaxis combined with quorum-sensing-like excitable chemical production and Vicsek-type polar alignment~\cite{ziepke2022multi}.

In this study, we introduce nematic interactions alongside chemotaxis, which gives rise to nematic order within the clusters. 
Within the simulated time window, nematic interactions enable cluster formation at lower chemotactic coefficients than in their absence, demonstrating that nematic interactions reinforce chemotaxis-driven aggregation.
By incorporating two distinct types of motility-based chemotactic responses—translational and rotational chemotaxis—we demonstrate that translational chemotaxis leads to dense, localized clusters concentrated at a single point, whereas rotational chemotaxis yields elongated, rod-like clusters.

Translational chemotaxis alters particle speed, trapping particles near the local maxima of chemical concentration generated by self-chemotaxis. These highly concentrated clusters are formed because the model lacks a mechanism that limits local particle density, such as excluded-volume interactions. Furthermore, because 
the alignment interaction produces strong orientational order, and the clusters generally retain finite polar order, the cluster itself translates at a finite velocity. In contrast, rotational chemotaxis leaves particle speed unchanged; thus, 
the aligned cluster continues to move at nearly the self-propulsion speed, leaving an elongated chemical trail.
Other particles are attracted to the resulting chemical trail, aligning along its direction upon arrival via nematic interactions. Because the trailing particles and the leading cluster move in nearly identical directions without approaching one another, a long, rod-like cluster is ultimately formed. Consequently, the inclusion of nematic interactions reveals novel cluster morphologies and internal ordering that are absent in the corresponding model without nematic interactions.

Both rotational and translational chemotaxis promote clustering, but through distinct mechanisms. Rotational chemotaxis has little effect on expanding the high-$\langle f_{\max}\rangle$ regime, yet it generally enhances $\langle f_{\max}\rangle$ values, leading to larger clusters.
Conversely, translational chemotaxis not only extends the high-$\langle f_{\max}\rangle$ domain to higher noise levels but also contributes to increasing $\langle f_{\max}\rangle$ itself.

Since the present model does not account for excluded volume interactions, excessive aggregation occurs. Incorporating excluded volume effects would facilitate a closer comparison with experimental studies using Janus rods, rod-like bacteria, and related systems.

%
\begin{acknowledgments}
The authors thank Gilhan Kim and Hiroki Ishikawa for  performing preliminary numerical simulations of related models.
\\
\end{acknowledgments}
%

\section*{DATA AVAILABILITY}

The data and simulation code that support the findings of this article are openly available in Zenodo~\cite{SakuraiZenodo2026}.

\bibliography{sakurai2026chemotaxis}
\end{document}